\documentclass[prl,aps,showpacs,onecolumn,floatfix]{revtex4}
\usepackage{epsfig} \usepackage{graphics} \usepackage{bm}
\usepackage{amssymb}
\usepackage{graphicx}
\begin{document}

\preprint{Natala-PRL}

\title{Non Fermi-Liquid
Magnetoresistance Oscillations in Quasi-One-Dimensional
Conductors}

\author{A.G. Lebed$^*$}

\affiliation{Department of Physics, University of Arizona, 1118 E.
4-th Street, Tucson, AZ 85721, USA}

\begin{abstract}
We theoretically demonstrate that strong non Fermi-liquid magnetic
oscillations of electron-electron scattering time can exist in
quasi-one-dimensional (Q1D) conductors under condition of the
magnetic breakdown between two open electron orbits. They are
shown to be due to electron-electron interactions in a metallic
phase under condition of the magnetic breakdown and they are
beyond the Fermi-liquid theory. In particular, we consider as
example the organic conductor (TMTSF)$_2$ClO$_4$ and perform both
analytical and numerical calculations for its known electron
spectrum. We also argue that similar oscillations of resistivity
can exist in a metallic phase of another Q1D organic conductor -
(Per)$_2$Au(mnt)$_2$.
\end{abstract}

\pacs{74.70.Kn}

\maketitle

Quasi-one-dimensional (Q1D) layered organic conductors exhibit
very rich and unusual properties in a magnetic field both in their
Field-Induced Spin(Charge)-Density-Wave (FIS(C)DW) and metallic
phases (for a review, see book [1]). Among them, the conductors
(TMTSF)$_2$ClO$_4$ and (TMTSF)$_2$PF$_6$ demonstrate the existence
of the so-called Rapid Magnetic Oscillations (RMO). It is
important that in (TMTSF)$_2$PF$_6$ the RMO exist only in FISDW
phase [2-5], whereas in the (TMTSF)$_2$ClO$_4$ they are observed
both in FISDW and metallic phases [6-11] (see recent Ref.[11] and
references therein). Yan et al. [7] first related the appearance
of the RMO in the (TMTSF)$_2$ClO$_4$ to the existence of the
so-called interference breakdown electron orbits [12,13] in the
(TMTSF)$_2$ClO$_4$ (see Fig.1). Theory of the RMO in the FISDW
phase was successfully created [14-17] by using the above
mentioned idea. Yan et al. [7] also related the experimentally
observed RMO in a metallic phase along ${\bf y}$ axis to one-body
effect - the so-called Stark interference between the interference
electron orbits (see Fig.1). On the other hand, it was stressed
[18], that the RMO in resistivity along the conducting chains in a
metallic phase of the organic conductor (TMTSF)$_2$ClO$_4$ cannot
be explained just by one-body magnetic breakdown through the
interference trajectories since the magnetic breakdown happened in
the perpendicular to the chain direction. It was suggested [18]
that non Fermi-liquid oscillations of electron-electron scattering
time under condition of the magnetic breakdown could, in
principal, explain the experimental observations. As to Q1D
conductor (TMTSF)$_2$PF$_6$, which doesn't exhibit the magnetic
breakdown, RMO in its FISDW have different physical origin and may
be explain in terms of the coexistence of two FISDW's [19,2-5]. It
is important that these second type of the RMO is also discovered
in FISDW phase of (TMTSF)$_2$ClO$_4$ [20].

The goal of our Letter is to make the exotic suggestion of
Ref.[18] more realistic and more suitable for its comparison with
the existing experiments [6-11] as well as for the possible future
experiments. First of all, here we consider the realistic Q1D
spectrum of the (TMTSF)$_2$ClO$_4$, instead of 2D spectrum of
Ref.[18]. In addition we perform numerical calculations of the
obtained results, instead of very rough estimations done in
Ref.[18]. Our conclusion is that non Fermi-liquid oscillations of
electron-electron scattering time can account for the RMO observed
in a metallic phase, although further experiments are needed. We
also suggest another candidate for discovery of non Fermi-liquid
oscillations of longitudinal resistivity under the condition of
the magnetic breakdown - Q1D organic conductor
(Per)$_2$Au(mnt)$_2$ under pressure in a metallic phase [21].

\begin{figure}[t]
\centering
\includegraphics[width=0.6\textwidth]{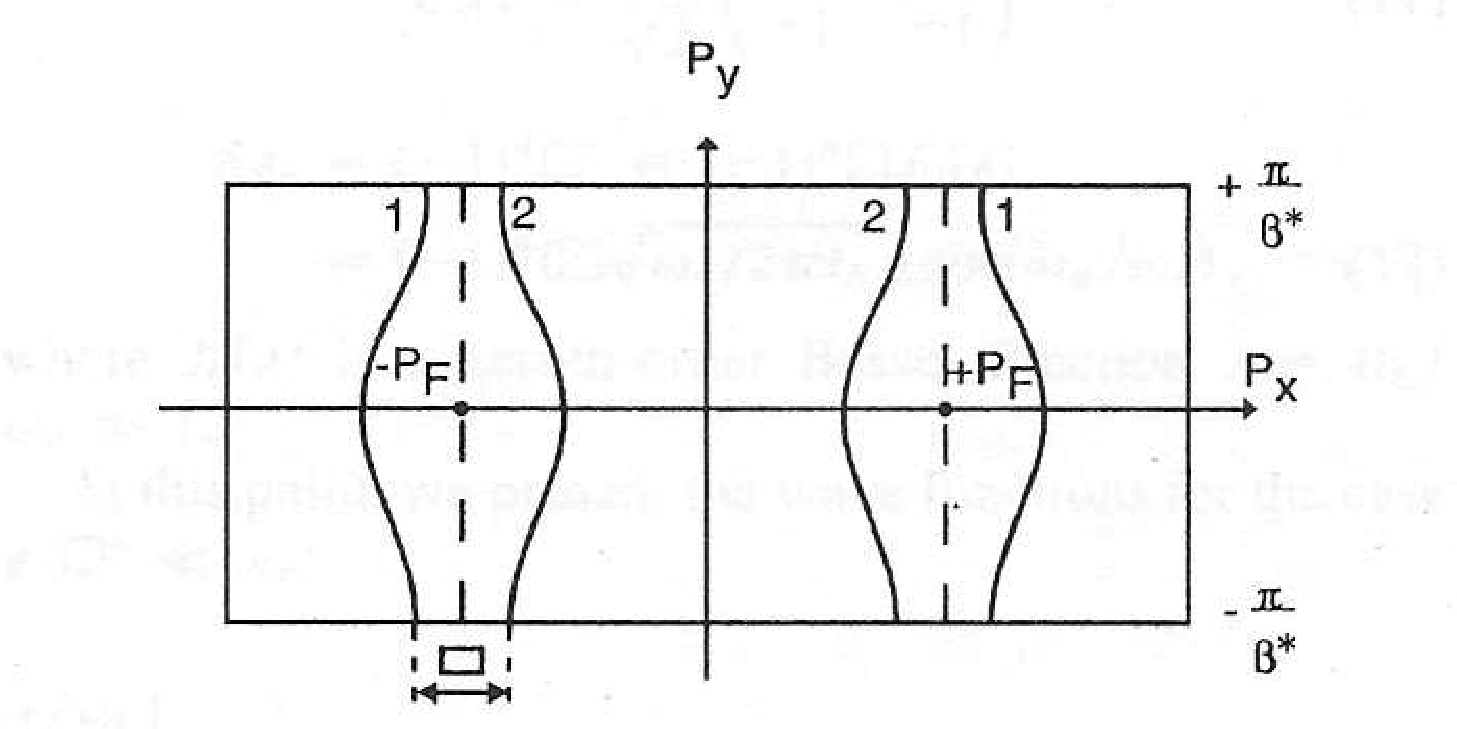}
\caption{Quasi-one-dimensional Fermi surface of the organic
conductor (TMTSF)$_2$ClO$_4$ in the presence of anion ordering
gap, $\Box \neq 0$ [see Eq.(5)].}
\end{figure}

Let us consider a typical Q1D electron spectrum of the organic
conductors (TMTSF)$_2$X (X=PF$_6$, ClO$_4$, AsF$_6$, etc.) in the
absence of the so-called anion ordering gap. It can be written in
tight-binding orthorhombic model as [1]
\begin{equation}
\epsilon^{\pm}({\bf p})= \pm v_F (p_x \mp p_F) + 2 t_b \cos(p_y
b^*) + 2 t_c \cos(p_z c^*),
\end{equation}
where $p_F \parallel x$, $b^* \parallel y$, and $c^* \parallel y$;
$v_F p_F \gg t_b \gg t_c$. The anion ordering gap in the conductor
(TMTSF)$_2$ClO$_4$, $\Box \ll 2t_b$, introduces the following
potential energy along ${\bf b^*}$ axis:
\begin{equation}
\Box(y)= \Box \cos(\pi y/b^*).
\end{equation}
It is possible to prove, using tight-binding approximation, that,
in the presence of the anion ordering gap potential (2), the
electron wave functions obey the following equations:
\begin{equation}
[\pm v_F (p_x \mp p_F) + 2t_b \cos(p_y
b^*)]\psi^{\pm}_{\epsilon}(p_y)
 + \Box \ \psi^{\pm}_{\epsilon}(p_y +\pi/b^*)=\epsilon \psi^{\pm}_{\epsilon}(p_y),
\end{equation}
\begin{equation}
[\pm v_F (p_x \mp p_F) - 2t_b \cos(p_y
b^*)]\psi^{\pm}_{\epsilon}(p_y+\pi/b^*)
 + \Box \ \psi^{\pm}_{\epsilon}(p_y)=\epsilon \psi^{\pm}_{\epsilon}(p_y+\pi/b^*),
\end{equation}
which double period along ${\bf b^*}$ axis and results in the
existing of the following four sheets of the Q1D Fermi surface:
\begin{equation}
\epsilon^{\pm}_{n}({\bf p})= \pm v_F (p_x \mp p_F) +(-1)^n
\sqrt{[2 t_b \cos(p_y b^*)]^2 + \Box^2} , \ \ n=1,2.
\end{equation}
[Note that here and below, to simplify the main equations, we
disregard energy dependence of the electron spectrum along ${\bf
z}$ axis, which we strictly take into account at the end of our
calculations.]
 It is important that the equation (5) corresponds to the
experimental situation in the conductor (TMTSF)$_2$ClO$_4$ at
ambient pressure.

In external magnetic field,
\begin{equation}
{\bf H} =(0,0,H), \ \ \ {\bf A}= (0,Hx,0),
\end{equation}
we perform the so-called Peierls substitutions [13,1]:
\begin{equation}
p_x \mp p_F \rightarrow - i \frac{d}{dx}, \ \ \ p_y \rightarrow
p_y - \frac{e}{c} A_y = p_y-\frac{e}{c} Hx.
\end{equation}
In this case, Eqs.(3) and (4) can be rewritten as

\begin{equation}
\biggl[ \mp i v_F \frac{d}{dx} + 2t_b \cos \biggl( p_y b^*
-\frac{\omega_b x}{v_F} \biggl)
\biggl]\psi^{\pm}_{\epsilon}(p_y,x)
 + \Box \ \psi^{\pm}_{\epsilon}(p_y +\pi/b^*,x)=\epsilon
 \psi^{\pm}_{\epsilon}(p_y,x),
\end{equation}

\begin{equation}
\biggl[ \mp i v_F \frac{d}{dx} - 2t_b \cos \biggl( p_y b^*
-\frac{\omega_b x}{v_F} \biggl)
\biggl]\psi^{\pm}_{\epsilon}(p_y+\pi/b^*,x)
 + \Box \ \psi^{\pm}_{\epsilon}(p_y,x)=\epsilon
 \psi^{\pm}_{\epsilon}(p_y+\pi/b^*,x),
\end{equation}
where $\omega_b=eH v_F b^*/c$ is the so-called cyclotron frequency
of electron motion along open electron trajectories in the
Brillouin zones [1]. Note that wave functions in Eqs.(8) and (9)
in the mixed representation are related to electron Bloch wave
functions as
\begin{equation}
\Psi^{\pm}_{\epsilon,p_y}(x,y)=\exp(ip_y
y)[\psi^{\pm}_{\epsilon}(p_y,x) +\exp(i \pi
y/b^*)\psi^{\pm}_{\epsilon}(p_y+\pi/b^*,x) ]
\end{equation}

At high magnetic fields, the magnetic breakdown phenomenon through
the anion gap, $\Box$, occurs between two open sheets of the Fermi
surfaces denoted by $n=1$ and $n=2$ in Eq.(5) (see Fig.1). The
corresponding magnetic breakdown field, $H_{MB}$, was calculated
in Ref.[16]:
\begin{equation}
H_{MB} = \frac{\pi c \Box^2}{2ev_F t_b b^*}.
\end{equation}
If we estimate experimental value of the field from measurements
[6-11], $H_{MB} \simeq 10-15 T$, we obtain from equation (11) that
$\Box \simeq 50K \ll 2t_b \simeq 400 K$. As known from a general
theory of the magnetic breakdown (see, for example, Ref.[9]), at
\begin{equation}
H \gg H_{MB} ,
\end{equation}
we can use theory of perturbation as it is done in
Refs.[14,15,17,18]. In this case, the first-order wave functions
are symmetrical and antisymmetrical combinations of two solutions
of Eqs.(8) and (9) at $\Box=0$ with the following corrections to
their energies [14]:
\begin{eqnarray}
&[\psi^{\pm}_1(p_y,x),\psi^{\pm}_1(p_y+\pi/b^*,x)]=\frac{\exp\biggl[
\pm i\frac{(\epsilon-\Box^*)x}{v_F} \biggl]}{\sqrt{2}}
&\biggl\{\exp \biggl[\pm \frac{i \lambda}{2} \sin
\biggl(p_yb^*-\frac{\omega_b x}{v_F} \bigg) \bigg],\exp
\biggl[\mp\frac{i \lambda}{2} \sin \biggl(p_yb^*-\frac{\omega_b
x}{v_F} \bigg) \bigg] \biggl\},
\end{eqnarray}

\begin{eqnarray}
[\psi^{\pm}_2(p_y,x),\psi^{\pm}_2(p_y+\pi/b^*,x)]=\frac{\exp
\biggl[ \pm i\frac{(\epsilon+\Box^*)x}{v_F}\biggl]}{\sqrt{2}}
\biggl\{\exp \biggl[\pm \frac{i \lambda}{2} \sin
\biggl(p_yb^*-\frac{\omega_b x}{v_F} \bigg) \bigg],-\exp
\biggl[\mp\frac{i \lambda}{2} \sin \biggl(p_yb^*-\frac{\omega_b
x}{v_F} \bigg) \bigg] \biggl\},
\end{eqnarray}
where
\begin{equation}
\lambda = \frac{4t_b}{\omega_b}.
\end{equation}
Note that in Eqs.(13) and (14) the electron energies are
\begin{equation}
\epsilon^{\pm}_1({\bf p})= \epsilon  - \Box^*, \ \ \epsilon = \pm
v_F(p_x \mp p_F),
\end{equation}
\begin{equation}
\epsilon^{\pm}_2({\bf p}) = \epsilon + \Box^*, \ \ \epsilon = \pm
v_F(p_x \mp p_F),
\end{equation}
where
\begin{equation}
\Box^* = J_0(\lambda) \Box \simeq \Box \ \sqrt{\frac{\omega_b}{2
\pi t_b}}\cos \biggl( \frac{4t_b c}{ev_F H b^*} \biggl) ,
\end{equation}
with $J_0(...)$ being the zeroth-order Bessel function. It is
important that energy levels (16) and (17) are oscillating
functions of an inverse magnetic field with the following period
(18):
\begin{equation}
\Delta \biggl( \frac{1}{H} \biggl) = \frac{\pi ev_F b^*}{4t_b c} .
\end{equation}

Let us discuss mechanism of conductivity along the conducting
chains. It is known that the considered conductors are very clean
[1],
\begin{equation}
\frac{1}{\tau} \simeq 0.1-1 \ K,
\end{equation}
therefore, the so-called electron-electron Umklapp scattering
processes [22,23],
\begin{equation}
{\bf p_1} + {\bf p_2} = {\bf p_3} + {\bf p_4} +4 p_F {\bf \hat
{x}},
\end{equation}
where ${\bf \hat {x}}$ is a unit vector along {\bf x} direction,
may play an important role and can define the in-chain resistivity
(see Fig.2).

\begin{figure}[t]
\centering
\includegraphics[width=0.55\textwidth]{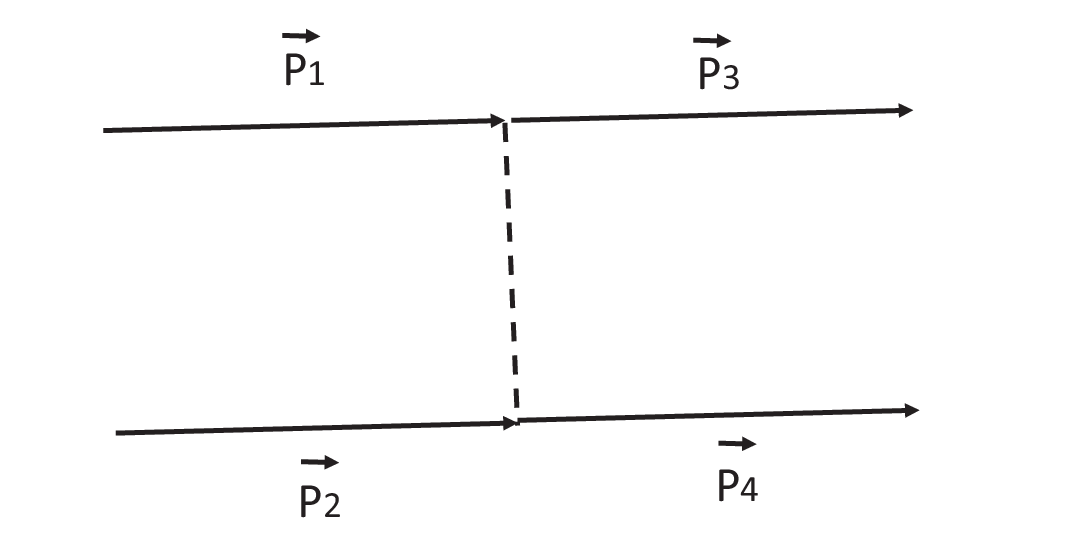}
\caption{Electron-electron scattering diagram, corresponding to
Umklapp process, ${\bf p_1+p_2 = p_3+p_4 }+4 p_F {\bf \hat {x}}$.
It defines the resistivity along conducting {\bf a} axes. }
\end{figure}
By means of variational principle for Boltzmann kinetic equation
for electron-electron scattering [18] and averaging probability of
Umklapp process (21), $U({\bf p_1},{\bf p_2};{\bf p_3},{\bf
p_4})$, by using Fermi-Dirac distribution functions,
$n[\epsilon({\bf p})]$, we obtain [18]:

 \begin{figure}[t]
\centering
\includegraphics[width=0.5\textwidth]{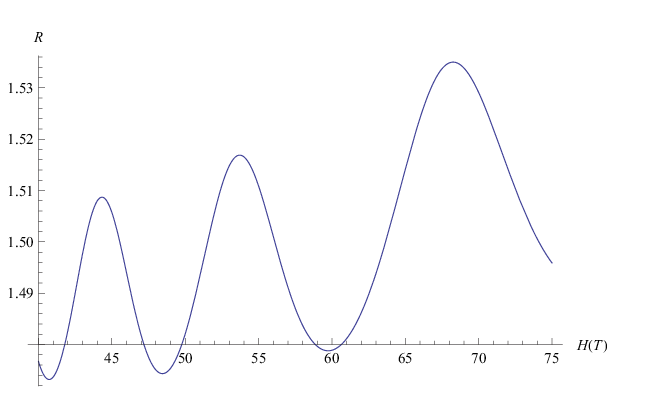}
\caption{Numerically calculated resistivity, including resistivity
oscillations, along the conduction chains is shown [see Eqs.(22)
and (23)]. }
\end{figure}

\begin{eqnarray}
\frac{1}{\tau} =\int_{-\infty}^{\infty}d^3p_1d^3p_2d^3p_3d^3p_4 \
U({\bf p_1},{\bf p_2};{\bf p_3},{\bf p_4})\delta({\bf p_1}+{\bf
p_2}-{\bf p_3}-{\bf p_4})
\nonumber\\
\times\delta[\epsilon({\bf p_1})+\epsilon({\bf p_2})-\epsilon({\bf
p_3})-\epsilon({\bf p_4})] n[\epsilon({\bf p_1})]n[\epsilon({\bf
p_2})](1-n[\epsilon({\bf p_3})])(1-n[\epsilon({\bf p_4})])
\end{eqnarray}
Generalizing Eq.(22) for the case, where there is the non-trivial
dependence of wave functions on coordinate $x$ in the magnetic
field, we find that
\begin{eqnarray}
\frac{1}{\tau}(H) = &&g^2 T \int^{\infty}_{- \infty} dx \frac{2
\pi T /v_F}{\sinh^2(2 \pi T x/v_F)} \biggl[\frac{2 \pi T
|x|/v_F}{\exp(4\pi T|x|/v_F)-1} + \frac{2\pi T |x|/v_F -1}{2}
\biggl]
\nonumber\\
&&\times \int^{\pi}_{-\pi}\frac{d\phi_1}{2\pi}
\int^{\pi}_{-\pi}\frac{d\phi_2}{2\pi} J^2_0[(4 t_c x/v_F)
\sin(\phi_1)]
\nonumber\\
 &&\times \biggl\{ J^2_0 \biggl[2 \lambda \sin \biggl(\frac{\omega_b x}{2v_F} \bigg)
\sin(\phi_2)\biggl] \cos^4 \biggl(\frac{\Box^* x}{v_F} \biggl) +
J^2_0 \biggl[2 \lambda \cos \biggl(\frac{\omega_b x}{2v_F} \bigg)
\cos(\phi_2)\biggl]  \sin^4 \biggl(\frac{\Box^*
x}{v_F}\biggl)\biggl\}
\end{eqnarray}
It is important that in Eq.(23) we take into account also free
electron motion along ${\bf z}$ axis [see Eq.(1)]. It is possible
to do taking the corresponding expression for motion along ${\bf
y}$ and put there $\Delta=0$ and $H \rightarrow 0$. Note that in
Q1D case resistivity along the chains
\begin{equation}
\rho (H) \sim \frac{1}{\tau}(H).
\end{equation}

We stress that in Ref.[18] the magnetoresistance was very roughly
estimated in the absence of $J^2_0[(4 t_c x/v_F) \sin(\phi_1)]$
term in Eq.(23) (i.e., in the absence of electron energy
dependence along ${\bf z}$ axis). Here we evaluate the entire
integral (23) numerically in the the interval of magnetic fields
of 40-75 T, which are much higher than the breakdown magnetic
field, experimentally estimated as 10-15 T [4-7]. For numeric
calculations of Eq.(23), we use the following values of the
parameters: $t_b = 200$K, $t_c=5$K, $\Box =50$K, $v_F=2 \times
10^7$ cm/sec, $b^*=7.7 \AA$, $c^*=13.6 \AA$ [1]. We perform
calculations for high enough temperature $T=10$K, which
corresponds to stabilization of a metallic phase in
(TMTSF)$_2$ClO$_4$ [6-11] (see Fig.3). As seen from Fig.3,
relatively large magnetoresistance oscillations,
\begin{equation}
\frac{\delta \rho}{\rho} \geq 10^{-2},
\end{equation}
can, indeed, exist in high magnetic fields in (TMTSF)$_2$ClO$_4$
organic conductor. The frequency of the calculated oscillations
can be estimated from Fig.3 as
\begin{equation}
\frac{H^2}{\Delta H} \simeq 250 \ T,
\end{equation}
which is very close to observed in (TMTSF)$_2$ClO$_4$ frequencies:
255T [6] and 265T [10]. It is logical to connect our current
theoretical results with the experimental RMO observed in its
metallic phase, although so far they have been studied at a little
bit lower magnetic fields. We suggest to investigate them
experimentally in high magnetic fields of the order of $H \simeq
50$ T to firmly reveal their non Fermi-liquid nature and to
quantitatively compare them with our calculations. The another
candidate for the experiments is layered Q1D conductor
(Per)$_2$Au(mnt)$_2$ [15].

The author is thankful to N.N. Bagmet (Lebed) for useful
discussions.

$^*$Also at: L.D. Landau Institute for Theoretical Physics, RAS, 2
Kosygina Street, Moscow 117334, Russia.

\end{document}